# Mean Interplanetary Magnetic Field Measurement Using the ARGO-YBJ Experiment

G. Aielli[1,2], C. Bacci[3,4], B. Bartoli[5,6], P. Bernardini[7,8], X.J. Bi[9], C. Bleve[7,8], P. Branchini[4], A. Budano[4], S. Bussino[3,4], A.K. Calabrese Melcarne[10], P. Camarri[1,2], Z. Cao[†9], A. Cappa[11,12], R. Cardarelli[2], S. Catalanotti[5,6], C. Cattaneo[13], P. Celio[13], S.Z. Chen[9], T.L. Chen[14], Y. Chen[9], P. Creti[7,8], S.W. Cui[15], B.Z. Dai[16], G. D'Alí Staiti[17,18], Danzengluobu[14], M. Dattoli[11,12,19], I. De Mitri[7,8], B. D'Ettorre Piazzoli[5,6], M. De Vincenzi[3,4], T. Di Girolamo[5,6], X.H. Ding[14], G. Di Sciascio[2], C.F. Feng[20], Z.Y. Feng[9], Zhenyong Feng[21], F. Galeazzi[4], P. Galeotti[11,19], R. Gargana[4], Q.B. Gou[9], Y.Q. Guo[9], H.H. He[9], Haibing Hu[14], Hongbo Hu[9], Q. Huang[21], M. Iacovacci[5,6], R. Iuppa[1,2], I. James[3,4], H.Y. Jia[21], Labaciren[14], H.J. Li[14], J.Y. Li[20], X.X. Li[9], B. Liberti[2], G. Liguori[13,22], C. Liu[9], C.Q. Liu[16], M.Y. Liu[15], J. Liu[16], H. Lu[9], X. H. Ma[9], G. Mancarella[7,8], S.M. Mari[3,4], G. Marsella[8,23], D. Martello[7,8], S. Mastroianni[5], X.R. Meng[14], P. Montini[3,4], C.C. Ning[14], A. Pagliaro[24,17], M. Panareo[8,23], L. Perrone[8,23], P. Pistilli[3,4], X.B. Qu[20], E. Rossi[5], F. Ruggieri[4], L. Saggese[6,5], P. Salvini[13], R. Santonico[1,2], P.R. Shen[9], X.D. Sheng[9], F. Shi[9], C. Stanescu[4], A. Surdo[8], Y.H. Tan[9], P. Vallania[11,12], S. Vernetto[11,12], C. Vigorito[11,12], B. Wang[9], H. Wang[9], C.Y. Wu[9], H.R. Wu[9], Z.G. Yao[9], B. Xu[21], L. Xue[20], Y.X. Yan[15], Q.Y. Yang[16], X.C. Yang[16], A.F. Yuan[14], M. Zha[9], H.M. Zhang[9], JiLong Zhang[9], JianLi Zhang[9], L. Zhang[16], P. Zhang[16], X.Y. Zhang[20], Y. Zhang[9], Zhaxisangzhu[14], Zhaxiciren[14], X.X. Zhou[21], F.R. Zhu[†21], Q.Q. Zhu[9], G. Zizzi[7,8] (The ARGO-YBJ COLLABORATION)

[1]Dipartimento di Fisica dell'Università "Roma Tor Vergata" - via della Ricerca Scientifica 1- 00133 Roma - Italy
[2]Istituto Nazionale di Fisica Nucleare - Sezione di Roma Tor Vergata - via della Ricerca Scientifica 1- 00133 Roma – Italy
[3]Dipartimento di Fisica dell'Università "Roma Tre" - via della Vasca Navale 84 – 00146 Roma – Italy
[4]Istituto Nazionale di Fisica Nucleare- Sezione di Roma3 - via della Vasca Navale 84 - 00146 Roma - Italy
[5]Istituto Nazionale di Fisica Nucleare - Sezione di Napoli - Complesso Universitario di Monte Sant'Angelo - via Cintia -80126 Napoli - Italy
[6] Dipartimento di Fisica dell'Università di Napoli - Complesso Universitario di Monte Sant'Angelo - via Cintia - 80126 Napoli – Italy
[7]Dipartimento di Fisica dell'Università del Salento - via per Arnesano - 73100 Lecce - Italy
[8]Istituto Nazionale di Fisica Nucleare - Sezione di Lecce - via per Arnesano -73100 Lecce - Italy
[9]Key Laboratory of Particle Astrophyics - Institute of High Energy Physics - Chinese Academy of Science - P.O. Box 918 - 100049 Beijing - China
[10] Istituto Nazionale di Fisica Nucleare - CNAF -viale Berti-Pichat 6/2 - 40127 Bologna – Italy
[11]Istituto Nazionale di Fisica Nucleare - Sezione di Torino - via P. Giuria 1 - 10125 Torino - Italy
[12]Istituto di Fisica dello Spazio Interplanetario dell'Istituto Nazionale di Astrofisica - corso Fiume 4 – 10133 Torino - Italy
[13]Istituto Nazionale di Fisica Nucleare - Sezione di Pavia - via Bassi 6 - 27100 Pavia – Italy
[14]Tibet University - 850000 Lhasa - Xizang - China,
[15]Hebei Normal University - Shijiazhuang 050016 - Hebei - China
[16]Yunnan University - 2 North Cuihu Rd – 650091 Kunming - Yunnan - China
[17]Istituto Nazionale di Fisica Nucleare - Sezione di Catania - Viale A. Doria 6 – 95125 Catania - Italy
[18] Università degli Studi di Palermo - Dipartimento di Fisica e Tecnologie Relative - Viale delle Scienze - Edificio 18 - 90128 Palermo - Italy
[19]Dipartimento di Fisica Generale dell'Università di Torino - via P. Giuria 1 - 10125 Torino -Italy
[20]Shandong University - 250100 Jinan -Shandong -.China,
[21]Southwest Jiaotong University - 610031 Chengdu - Sichuan - China
[22]Dipartimento di Fisica Nucleare e Teorica dell'Università di Pavia - via Bassi 6 - 27100 Pavia - Italy
[23]Dipartimento di In gegneria dell'Innovazione -Università del Salento - 73100 Lecce - Italy
[24]Istituto di Astrofisica Spaziale e Fisica Cosmica di Palermo - Istituto Nazionale di Astrofisica - via Ugo La Malfa 153 - 90146 Palermo – Italy

**Abstract**

The sun blocks cosmic ray particles from outside the solar system, forming a detectable shadow in the sky map of cosmic rays detected by the ARGO-YBJ experiment in Tibet. Because the cosmic ray particles are positive charged, the magnetic field between the sun and the earth deflects them from straight trajectories and results in a shift of the shadow from the true location of the sun. Here we show that the shift measures the intensity of the field which is transported by the solar wind from the sun to the earth.

***Key-words:*** Cosmic Ray, Solar Wind, Magnetic Field,

[†]correspondence authors, E-mail addresses: zhufr@ihep.ac.cn and caozh@ihep.ac.cn

## 1. Introduction

Cosmic rays from outside the solar system, mainly hydrogen and helium nuclei (Ahn *et al.*, 2010), isotropically arrive at the earth and can be recorded by detectors on the ground, such as the Resistive Plate Chamber (RPC) array in the ARGO-YBJ Experiment (Aielli *et al.*, 2006) at 4300 m above sea level in Tibet, China. At energies around 5TeV, the cosmic ray arrival directions are measured by the ARGO-YBJ detector with accuracy better than 1° (Iuppa *et al.*, 2009; Aielli *et al.*, 2010). The distribution of particle counts on the sky shows a deficit corresponding to the location of the sun's shadow. The magnetic fields distributed in the interplanetary space (abbreviated as IMF) along the trajectories deflect these rays slightly and shift the shadow from the true location of the sun. The Tibet AS$_\gamma$ experiment observed the effect for the first time (Amenomori *et al.*, 1993). With its sensitivity, year-round observation for the Tibet AS$_\gamma$ experiment was required to make a significant sun shadow map, therefore only the yearly variation of the shadow was reported (Amenomori *et al.*, 2006). As a part of the magnetic field on the photosphere spread out to the interplanetary space by the solar wind (Severny *et al.*, 1970; Wilcox, 1971), IMF is usually monitored using orbiting detectors only at a distance of approximately 1 *AU* from the sun. Abnormally strong fluctuations of IMF could severely disturb the geomagnetic environment. What is measured using the deflection of cosmic rays is a cumulative effect along the whole path from the sun to the earth. Since it is strongly modulated by solar activities, IMF is better studied in quiet phases of the sun. The observation using the ARGO-YBJ experiment described in this paper was made just in such a particularly good time window when the solar activity had stayed at its minimum for an unexpectedly long time since 2006.

At distances greater than 5 solar radii from the sun center, the IMF is distributed mainly in the ecliptic plane (Ness & Wilcox 1965, Balogh & Jokipii 2009)). Its z-component perpendicular to the ecliptic plane deflects cosmic rays in the east-west direction, therefore it drives the shadow with an extra shift in addition to the geomagnetic effect which constantly moves the sun shadow towards west as observed in the moon shadow measurement (Iuppa *et al.*, 2009). Its y-component, $B_y$, in the ecliptic plane defined to be perpendicular to the line of sight, deflects cosmic rays and thus drives the sun shadow in the north-south direction. It has no contamination from the geomagnetic effect because the declination angle of the geomagnetic field is less than 0.5° at the ARGO-YBJ site. The measurement of $B_y$ is the topic of this paper.

## 2. The ARGO-YBJ Experiment Data and the Sun Shadow Analysis

The ARGO-YBJ detector is more sensitive than the Tibet AS$_\gamma$ experiment with its fully covered active area of 5800 m$^2$. This allows an investigation for variations of the shadow in shorter periods compared with previous observations. New phenomena associated with both spatial patterns and temporal behaviours of IMF can be observed with the ARGO-YBJ data from July 2006 to October 2009, namely 903 exposure days in total. Well reconstructed events from directions within 6° around the sun are selected. They must be also within 150m from the centre of the array and fire at least 100 56×62 cm2 spatial pixels of the detector. These criteria guarantee the event reconstruction quality. 86 million events survive the selection and are used for the study described here. Using this data set, a map of the sun shadow is plotted in Fig.1, with the most significant point (45$\sigma$) located at (0.17±0.02)° toward north and (0.26±0.02)° toward west. In order to understand the systematic errors of the measurement, the same selection criteria are applied to events around the moon. With a significance of 55$\sigma$, similar systematic shifts of the moon shadow location are observed. The reason of the (0.19±0.02)° shift toward north is under investigation while that of (0.31±0.02)° towards west is exactly expected due to the geomagnetic effect (Iuppa *et al.*, 2009).

In order to study the spatial distribution of IMF over solar longitudes, the ARGO-YBJ data mentioned above are divided into 12 groups according to the position of the earth in terms of solar longitudes when the events are recorded. More specifically, events in each group fly along trajectories within a sector of 30° in the ecliptic. In order to compensate for the earth orbital effect, the synodic Carrington period of 27.3 days and corresponding Carrington longitudes are used to describe the position of the earth in the sectors. Rotating together with the sun, IMF is nearly frozen in the sector within the 8 minutes needed for the cosmic rays to fly over the distance of 1 *AU* and be deflected as well. Each group has approximately 7 million events which measure the shadow with an average significance of about 10$\sigma$. The position of the shadow is measured by projecting the 2-dimensional map onto the axis along the north-south direction and fitting it with a Gaussian functional form. In this way, the position is measured with an accuracy of 0.02°. A clear shift of the sun shadow over a range greater than 0.8° from the southernmost to the northernmost position is observed as plotted in both panels of Fig.2. This clearly reveals periodical distribution patterns over solar longitudes, indicating that cosmic rays are deflected differently by IMF in the 12 sectors. The angular position of the shadow can be used as a measure of IMF as a function of the solar longitude. Since only the shadow displacement is relevant to the topic of this paper, the systematic offset on its location of 0.19°, which is found in the moon shadow observation, is removed in Fig.2.

In the upper panel of Fig.2, we used the ARGO-YBJ events collected from January 2008 to April 2009. It is found that the shadow reaches its northernmost position at about 170°, then switches towards south and monotonously reaches its south-most position at about 320°, and subsequently switches again towards north. This indicates that the polarization of the y-component of the field oscillates correspondingly. It follows that the field is spatially distributed in a bisector pattern which is also observed by the satellite-borne detectors as shown in the corresponding data which is downloaded from (King & Papitashvili, 2004). This motivates the classification of the data into two groups, i.e. $G_1$, which is the one under discussion, and $G_2$, which includes all the other events. Using the x-component along the direction connecting the sun and the observer, the satellites measurement of the magnetic fields clearly separates the adjacent sectors according to the polarization which points toward the sun or away.

It is very interesting to observe a sudden switch from a 4-sector to a simpler bisector pattern in July 2007 and a subtler return to the 4-sector pattern around April 2009 with data from (King & Papitashvili, 2004). Using the ARGO-YBJ data in $G_2$, the observed displacement of the sun shadow is found varying in a similar 4-sector pattern as shown in the lower panel of Fig.2. Between the north-most position at about 80° and the south-most one at about 330°, the shadow switches its drifting direction back and forth again between 160° and 260°, as in a sub-period. For both cases, the spatial patterns in terms of the longitudinal distribution of the sun shadow displacements are well fitted using harmonic functional forms as shown in Fig.2 (solid curves) and the parameters are displayed in the legends of the figures.

### 3. IMF Measurement and Discussion

Based on this observation, one can estimate the transverse component of the IMF, $B_y$, by fitting the measured displacement distribution as a function of the solar longitude with a minimal assumption in its model (Parker, 1963; Amenomori et al., 2000). This model has been thoroughly described in (Amenomori et al., 2000). The major difference is that the field has a longitudinal distribution $f(\psi)$ that is assumed to be the same as the distribution of the displacement of the observed sun shadow with a phase shift $\delta$. The model is briefly summarized as follows. IMF is distributed only in the ecliptic for $r > 5 R_{sun}$, where $r$ is the distance from the centre of the sun and $R_{sun}$ the solar radius. The distribution on solar longitudes is conserved along spiral trajectories of the solar wind starting at points on the photosphere surface. The projecting direction on the photosphere is determined by the solar wind speed, $v_r$ (perpendicular to the surface), together with the rotating speed of the sun, $r\omega$, where $\omega$ is the angular speed of the sun rotation. The intensity of IMF falls as $1/r$ along the spirals at distance far from the sun. That is

$B_x = B_0 (R_{sun}/r)^2 f(\psi - \delta)$

$B_y = B_0 (R_{sun}/r)^2 (r\omega/v_r) f(\psi - \delta)$,

where $B_0$ is the field intensity at the solar surface. $B_0$ and $\delta$ are the two parameters that can be estimated with our data. The solar wind that transports the fields away from the sun has a $v_r$ distributed over a range from 290 to 700 km/s. For simplicity, we assume the average value of 400 km/s in the analysis (King & Papitashvili, 2004).

For cosmic rays around 5 TeV, the composition has been well measured in balloon experiments (Ahn et al., 2010). A ray-tracing algorithm is developed based on this IMF model to describe particles passing by the sun. The ray-tracing is carried out with the known composition in order to reproduce the shadow effect due to the screening of the sun. It is found that the cosmic rays are less deflected as they fly farther from the sun because the IMF intensity decreases as $1/r$; however, at these larger distances there is a larger shift of the sun shadow position due to a perspective effect.

For instance, approximately 73% of the shift of the shadow formed by 1TeV cosmic rays occurs in the second half of their journey from the sun to the earth. Using this ray-tracing tool, it is straightforward to figure out the intensity of IMF $B_0$ and the phase shift $\delta$ by fitting the sun shadow displacement data shown in Fig.2. The calculated displacements are shown in the same figure as filled squares.

The solar wind needs about 4.5 days to transport IMF from the photosphere surface to the earth where the field intensity is measured by the orbiting detectors. On the other hand, cosmic ray particles take only 8 minutes to fly over the "frozen" IMF, thus the measurement by using the deflection of cosmic rays is an average over the fields along different spirals that start from different longitude regions on the photosphere surface. This is equivalent to an average over many reference points along the orbit of the earth in a range ahead of the current earth position in terms of Carrington solar longitudes. It results in a phase shift between the sun shadow displacements and the IMF distribution in solar longitudes. According to the ray-tracing simulation, the phase shift is calculated to be 21° corresponding to 1.6 days ahead. Applying this phase shift to the IMF intensities at $r = 1$ *AU* measured by the sun shadow displacements, we obtain measurements of $B_y$ as plotted in the upper and lower panels of Fig. 3 for periods $G_1$ and $G_2$, respectively. The solid curves, with uncertainty represented by the shaded area, are the results of the ARGO-YBJ experiment while the solid dots represent the measurements by the orbiting detectors (King & Papitashvili, 2004). The two measurements are of the same order in the amplitude of $(2.0\pm0.2)$ nT and are consistent in the alternating periodical pattern. In order to carry out the observation in a sector of 30°, the current ARGO-YBJ experiment must be operated for approximately 13 complete Carrington periods to collect enough events for the required accuracy on the position of the sun shadow, namely 0.1°, in each sector. The same folding technique with a Carrington period is used in the satellite detector data analysis as mentioned.

The measurement is particularly interesting because it refers to a "ground state" of IMF in the quiet phase, around the minimum between the 23rd and the 24th solar cycles. It is also an important complement to the regular monitoring of IMF at satellite orbits because it is related to the IMF distribution over the much larger scale of $0.5 AU$.

Even more importantly, this measurement could foresee fluctuations of IMF which will sweep the earth about 2 days later, demonstrating a potential forecasting capability for magnetic storms due to solar events. Although the ARGO-YBJ detector has not sufficient sensitivity to measure the sun shadow with the required precision in a single day, a future detector such as the LHAASO project (Cao, 2010) may be sufficient to be able to measure the sun shadow within one day or so. Thus it would be practically useful in monitoring unexpectedly large shifts of the sun shadow to foresee magnetic storms due to solar events approximately two days ahead of their arrival to the earth. This triggers further study on a possible new method for space weather monitoring among many other well developed methods (Kudela, 2007). For instance, the z-component of IMF, which may be even more relevant to the potential forecasting, is under analysis with a careful decoupling from geomagnetic effects.

**Acknowledgments** This work is supported in China by NSFC (No.10120130794), the Chinese Ministry of Science and Technology, the Chinese Academy of Sciences, the Key Laboratory of Particle Astrophysics, CAS, and in Italy by the Istituto Nazionale di Fisica Nucleare (INFN). We also acknowledge the essential supports of W.Y. Chen, G. Yang, X.F. Yuan, C.Y. Zhao, R.Assiro, B.Biondo, S.Bricola, F.Budano, A.Corvaglia, B.D'Aquino, R.Esposito, A.Innocente, A.Mangano, E.Pastori, C.Pinto, E.Reali, F.Taurino and A.Zerbini, in the installation, debugging and maintenance of the detector.

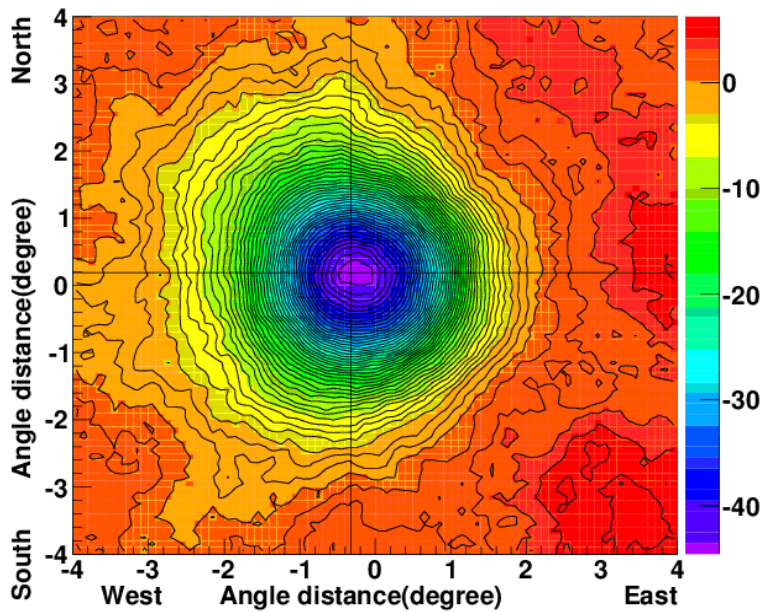

**Fig.1.** The sun shadow measured using all the data taken by the ARGO-YBJ experiment from July 2006 to October 2009. This map is based on a 2-D histogram on a grid of 0.1° by 0.1° with a smearing within a circular bin of 1.2° as its angular radius which takes into account the point-spread function of the ARGO-YBJ detector. The central circle of the contour map indicates a significance of -43.6 standard deviation of the deficit in event count. The step between contour lines is one standard deviation. The maximum significance is -44.6 standard deviation near the cross which marks the most significant location of the sun shadow. $3.253 \times 10^6$ events are found in the central circular bin as the on-source event counting, while the expected background is $3.34 \times 10^6$.

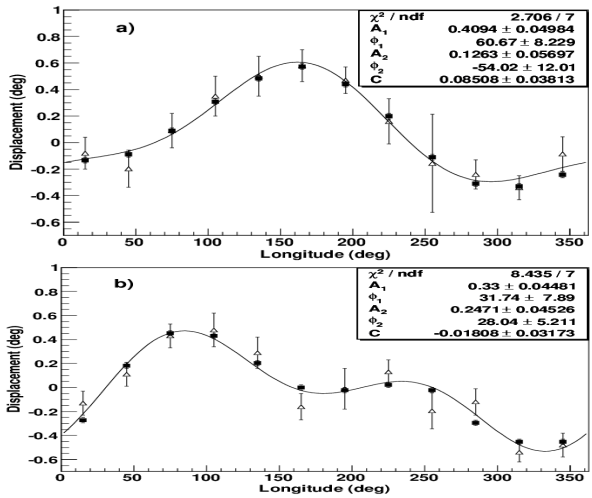

**Fig.2.** Shift of the centre of the sun shadow along the north-south direction during a complete Carrington period. The horizontal axis gives the Carrington longitude and the vertical axis is the angular displacement of the centre of the shadow. In the upper panel a), the observation (triangles) in the period $G_1$ reveals that the shadow walks towards north in nearly half of the Carrington period and towards south in the rest of the period. The curve is a fit with a harmonic functional form. The squares represent the displacements of the calculated sun shadows. In the lower panel b), a similar shift of the shadow but with different pattern is observed in period $G_2$, i.e. the shadow moves from side to side twice per Carrington period. $C$, $A_i$ and $\phi_i$ ($i=1,2$) are parameters (0-th and i-th order coefficients and phase shifts) of the second order harmonic functional form.

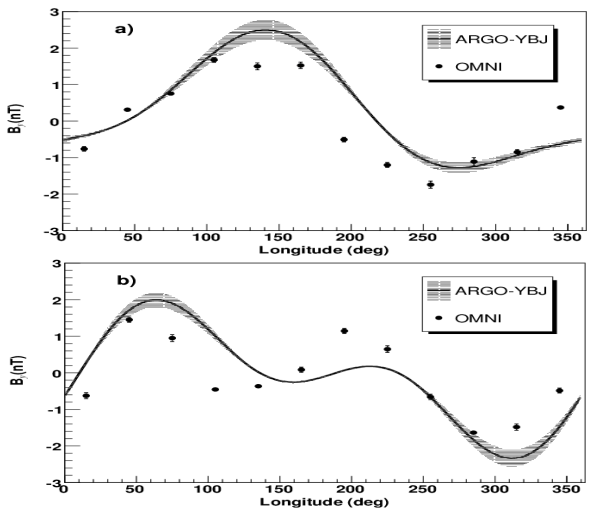

**Fig.3.** Comparison between two measurements of IMF using different methods. The solid curve represents the field component $B_y$ near the earth measured by the ARGO-YBJ experiment using cosmic ray deflection. In period $G_1$ (upper panel a)) a clear bisector pattern is observed. Positive sign indicates that the field is pointing to the centre of the sun. An uncertainty of one standard deviation is marked by the shaded area. The solid dots represent the measurements using the OMNI observational data downloaded from (King & Papitashvili, 2004). In the lower panel b), the results with the 4-sector structure in period $G_2$ are displayed.